\RequirePackage{ifpdf}
\ifpdf 
\documentclass[pdftex]{sigma}
\else
\documentclass{sigma}
\fi

\numberwithin{equation}{section}

\usepackage{bbm}
\usepackage{float}
\usepackage{wrapfig}
\DeclareMathAlphabet{\boldmathe}{T1}{cmr}{bx}{it} \font\af=msbm10

\begin{document}

\allowdisplaybreaks

\renewcommand{\PaperNumber}{006}

\FirstPageHeading

\renewcommand{\thefootnote}{$\star$}

\newcommand{\refs}[1]{(\ref{#1})}
\newcommand{\mbf}[1]{\boldmathe{#1}}
\def\mtxt#1{\quad\hbox{{#1}}\quad}
\def\mbx{\mbf{x}}
\def\mby{\mbf{y}}
\def\tr{\hbox{tr}}
\def\id{\mathbbm{1}}
\def\Lx{L_{\mbx}}
\def\Ly{L_{\mby}}
\def\Seff{S_{\rm eff}}
\def\gam{\gamma}
\def\lam{\lambda}
\def\Z{\mbox{\af Z}}

\ShortArticleName{Generalized Potts-Models and their Relevance for
Gauge Theories}

\ArticleName{Generalized Potts-Models and their Relevance\\ for
Gauge Theories\footnote{This paper is a contribution to the
Proceedings of the O'Raifeartaigh Symposium on Non-Perturbative
and Symmetry Methods in Field Theory
 (June 22--24, 2006, Budapest, Hungary).
The full collection is available at
\href{http://www.emis.de/journals/SIGMA/LOR2006.html}{http://www.emis.de/journals/SIGMA/LOR2006.html}}}

\Author{Andreas WIPF~$^\dag$, Thomas HEINZL~$^\ddag$, Tobias
KAESTNER~$^\dag$ and Christian WOZAR~$^\dag$}
\AuthorNameForHeading{A. Wipf, T. Heinzl, T. Kaestner and C. Wozar}

\Address{$^\dag$~Theoretisch-Physikalisches Institut,
Friedrich-Schiller-University Jena, Germany}

\EmailD{\href{mailto:wipf@tpi.uni-jena.de}{wipf@tpi.uni-jena.de}}
\URLaddressD{\url{http://www.personal.uni-jena.de/~p5anwi/}}
\Address{$^\ddag$~School of Mathematics and Statistics, University
of Plymouth, United Kingdom}
\EmailD{\href{mailto:thomas.heinzl@plymouth.ac.uk}{thomas.heinzl@plymouth.ac.uk}}

\ArticleDates{Received October 05, 2006, in f\/inal form December
12, 2006; Published online January 05, 2007}

\Abstract{We study the Polyakov loop dynamics originating from
f\/inite-temperature Yang--Mills theory. The ef\/fective actions
contain center-symmetric terms involving powers of the Polyakov
loop, each with its own coupling. For a subclass with two
couplings we perform a detailed analysis of the statistical
mechanics involved. To this end we employ a modif\/ied mean
f\/ield approximation and Monte Carlo simulations based on a novel
cluster algorithm. We f\/ind excellent agreement of both
approaches. The phase diagram exhibits both f\/irst and second
order transitions between symmetric, ferromagnetic and
antiferromagnetic phases with phase boundaries merging at three
tricritical points. The critical exponents $\nu$ and $\gamma$ at
the continuous transition between symmetric and antiferromagnetic
phases are the same as for the 3-state spin Potts model.}

\Keywords{gauge theories; Potts models; Polyakov loop dynamics;
mean f\/ield approximation; Monte Carlo simulations}

\Classification{81T10; 81T25; 81T80}

\section{Introduction}

Symmetry constraints and strong coupling expansion for the
ef\/fective action describing the Polyakov loop dynamics of gauge
theories lead to ef\/fective f\/ield theories with rich phase
structures. The f\/ields are the fundamental characters of the
gauge group with the fundamental domain as target space. The
center symmetry of pure gauge theory remains a symmetry of the
ef\/fective models. If one further freezes the Polyakov loop to
the center ${\cal Z}$ of the gauge group one obtains the well
known vector Potts spin-models, sometimes called clock models.
Hence we call the ef\/fective theories for the Polyakov loop
dynamics \emph{generalized ${\cal Z}$-Potts models}. We review our
recent results on generalized $\Z_3$-Potts models \cite{wozar}.
These results were obtained with the help of an improved mean
f\/ield approximation and Monte Carlo simulations. The mean
f\/ield approximation turns out to be much better than expected.
Probably this is due to the existence of tricritical points in the
ef\/fective theories. There exist four distinct phases and
transitions of f\/irst and second order. The critical exponents
$\nu$ and $\gamma$ at the second order transition from the
symmetric to antiferromagnetic phase for the generalized Potts
model are the same as for the corresponding Potts spin model.

Earlier on it had been conjectured that the ef\/fective Polyakov
loop dynamics for f\/inite tempe\-rature $SU(N)$ gauge theories
near the phase transition point is very well modelled by
$3$-di\-mensional $\Z_N$ spin systems
\cite{Yaffe:1982qf,Bazavov:2006ny}. For $SU(2)$ this conjecture is
supported by universality arguments and numerical simulations. The
status of the conjecture for $SU(3)$ gauge theories is unclear,
since the phase transition is f\/irst order such that universality
arguments apparently are not applicable.

\renewcommand{\thefootnote}{\arabic{footnote}}
\setcounter{footnote}{0}

\section{Recall of planar Potts models}\label{sectpotts}
\begin{wrapfigure}[14]{l}{7cm}
\centerline{\includegraphics{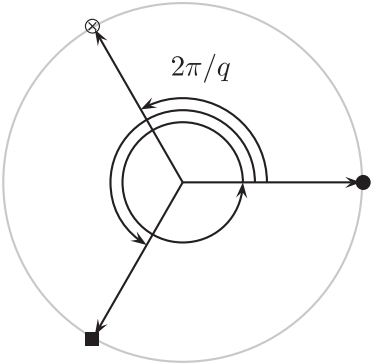}}
\caption{\label{fig:angles} Angles in a $q$-state Potts model.}
\end{wrapfigure}

The $q$-state Potts model
\cite{pottsmodel,pottsmodel1,pottsmodel2} is a natural extension
of the Ising model.  On every lattice site~$x$ there is a planar
vector with unit length which may point in $q$ dif\/ferent
directions (Fig.~\ref{fig:angles}) with ang\-les
\begin{equation*}
\theta_x\in\left\{\frac{2\pi}{q},\frac{4\pi}{q},\dots,2\pi\right\}=\Z_q.
\end{equation*}
Only nearest neighbors interact and their contribution to the
energy is proportional to the scalar product of the vectors, such
that
\begin{equation}
H=-J \sum_{\langle
xy\rangle}\cos\left(\theta_x-\theta_y\right).\label{ham}
\end{equation}
The Hamiltonian $H$ is invariant under simultaneous rotations of
all vectors by a multiple of $2\pi/q$. These $\Z_q$ symmetries map
a conf\/iguration $w=\{\theta_x\vert x\in\Lambda\}$ into
$w'=\{\theta_x+2\pi n/q\},\; n\in \{1,\dots,q\}$.
In two and higher dimensions the spin model shows a~phase
transition at a critical coupling $K_c=\beta J_c>0$ from the
symmetric to the ferromagnetic phase. In two dimensions this
transition is second order for $q\leq 4$ and f\/irst order for
$q>4$. In three dimensions it is second order
for $q\leq 2$ and f\/irst order for $q>2$.

\begin{wrapfigure}[15]{l}{7cm}
\centerline{\includegraphics{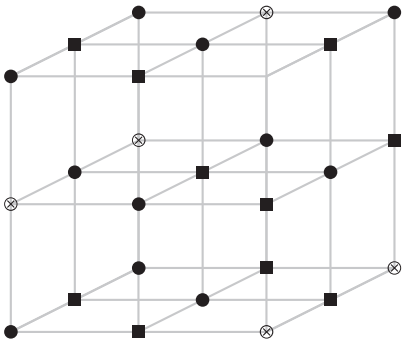}}
\caption{\label{fig:coloring} Coloring of neighboring lattice
sites.}
\end{wrapfigure}
There is another phase transition at negative cri\-ti\-cal
coupling $K'_{c}$ from the symmetric to the antiferromagnetic
phase. For $q\geq 3$ and negative $K$ the number of ground states
of \refs{ham} increases rapidly with the number of lattice sites.
It equals the number of ways  one can color the vertices with $q$
colors such that two neighboring sites have dif\/ferent colors
(Fig.~\ref{fig:coloring}). In the antiferromagnetic case the
dege\-ne\-rate ground states contribute considerably to the
\emph{entropy} \cite{pottsaf}
\begin{equation*}
S_B(P)=-\sum P(w)\log P(w),
\end{equation*}
where one sums over all spin conf\/igurations and $P$ is the
probability of $w$. We use the well known varia\-tio\-nal
characterization of the \emph{free energy},
\begin{equation*}
\beta F=\inf_{P} \left(\beta \langle H\rangle_P-S_B\right),\qquad
\langle H\rangle_P=\sum_w P(w)H(w),
\end{equation*}
where the minimum is to be taken on the space of all probability
measures. The unique minimizing probability measure is the Gibbs
state
\begin{equation}
 P_{\rm Gibbs}\sim e^{-\beta H}\label{gibbsstate}
\end{equation}
belonging to the canonical ensemble. In the variational
def\/inition of the \emph{convex effective action} one minimizes
on the convex subspace of probability measures with f\/ixed mean
f\/ield,
\begin{equation}
\Gamma[m]=\inf_{P}\big(\beta \langle H\rangle_P -S(P)\,\big\vert
\langle e^{i\theta_x}\rangle_P =m(x)\big).\label{effaction}
\end{equation}
The f\/ield $m(x)$ which minimizes the ef\/fective action is by
construction the expectation value of the f\/ield $e^{i\theta_x}$
in the thermodynamic equilibrium state~\refs{gibbsstate}.

In the mean f\/ield approximation to the ef\/fective action one
further assumes that the measure is a product measure
\cite{meanfield,meanfield1},
\begin{equation}
P(w)=P\left(\{\theta_x\}\right)=\prod_x
p_x(\theta_x),\label{productstate}
\end{equation}
where the single site probabilities are maps $p_x:\Z_q\to [0,1]$.
The approximate ef\/fective action is denoted by $\Gamma_{\rm
mf}[m]$.

The symmetric and ferromagnetic phases are both translationally
invariant. In the mean f\/ield approximation all single site
probabilities are the same, $p_x=p$, and the mean f\/ield is
constant, $m(x)=m$. The \emph{effective potential} is the
ef\/fective action for constant mean f\/ield, divided by the
number of sites. Its mean f\/ield approximation is
\begin{equation}
u_{\rm mf}(m)=\inf_{p}\left(-K mm^*+\sum_\theta p(\theta)\log
p(\theta)\Big\vert \sum_\theta  p
(\theta)\,e^{i\theta}=m\right),\qquad K=dJ.\label{effpotmf}
\end{equation}
It agrees with the mean f\/ield approximation to the constraint
ef\/fective potential, introduced by O'Raifeartaigh et
al.~\cite{ceffpot,ceffpotmf}. In the antiferromagnetic phase there
is no translational invariance on the whole lattice $\Lambda$ but
on each of two sublattices in the decomposition
$\Lambda=\Lambda_1\cup\Lambda_2$. The sublattices are such that
two nearest neighbors always belong to dif\/ferent sublattices.
Thus the single site distributions $p_x$ in \refs{productstate}
are not equal on the whole lattice, but only on the sublattices,
\begin{equation}
 p_x=p_1\ \hbox{ on } \ \Lambda_1 \quad \mtxt{and}\quad
p_x=p_2\ \hbox{ on } \ \Lambda_2.\label{prodstate2}
\end{equation}
The minimization of the ef\/fective action on such product states
is subject to the constraints
\begin{equation*}
\sum_{\theta\in\,\Z_q} p_1(\theta) e^{i\theta}=m_1\quad
\mtxt{and}\quad
\sum_{\theta\in\, \Z_q} p_2(\theta) e^{i\theta}=m_2
\end{equation*}
and yields the following mean f\/ield ef\/fective potential
\begin{equation*}
u_{\rm mf}\left(m_1,m_2\right)=\frac{1}{2}\left(K\vert
m_1-m_2\vert^2
+\sum_i u_{\rm mf}(m_i)\right),
\end{equation*}
where $u_{\rm mf}$ is the ef\/fective potential \refs{effpotmf}.
For $K>0$ the minimum is attained for $m_1=m_2$ and translational
invariance is restored. In the symmetric and ferromagnetic phases
the single site probabilities $p_1=p_2=p$. In the symmetric phase
$p=1/q$ for every orientation and in the ferromagnetic phase $p$
is peaked at one orientation. Hence there exist $q$ dif\/ferent
ferromagnetic equilibrium states related by $\Z_q$ symmetry
transformations. In the antiferromagnetic phase the probabilities
$p_1$ and $p_2$ are dif\/ferent.

\begin{figure}[ht]
\centerline{\includegraphics{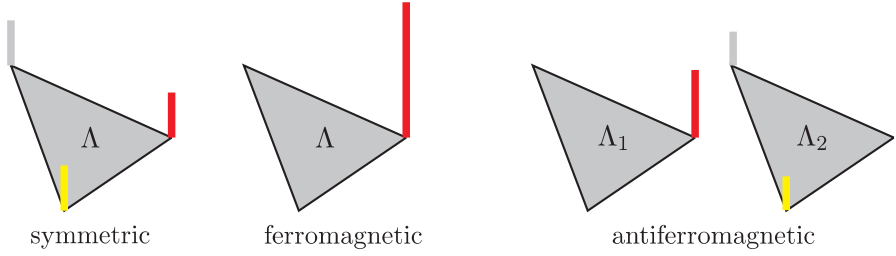}}
\caption{\label{fig:distributions} Single site distributions of
the $3$-state Potts model.}
\end{figure}

For the $3$-state Potts model  one can calculate the single site
probabilities explicitly. On one sublattice it is peaked at one
orientation and on the other sublattice it is equally distributed
over the remaining $2$ orientations. Hence there are $6$
dif\/ferent antiferromagnetic equilibrium states related by $\Z_3$
symmetries and an exchange of the sublattices. The results for
single site distributions of the $3$-state Potts model are
depicted in Fig.~\ref{fig:distributions}.

\section{Polyakov-loop dynamics}\label{sectpolyakov}

We consider pure Euclidean gauge theories with group valued link
variables $U_{x;\mu}$ on a lattice with $N_t$ sites in the
temporal direction. The f\/ields are periodic in this direction,
$U_{t+N_t,\mbx;\mu}=U_{t,\mbx;\mu}$. We are interested in the
distribution and expectation values of the traced Polyakov loop
va\-riable
\begin{gather}
\Lx=\tr \,{\cal P}_\mbx,\qquad {\cal P}_{\mbx}=\prod_{t=1}^{N_t}
U_{t,\mbx;0},\label{constr}
\end{gather}
since $\langle L_\mbx\rangle$ is an order parameter for f\/inite
temperature gluodynamics (see \cite{holland} for a review). In the
low-temperature conf\/ined phase $\langle \Lx\rangle=0$ and in the
high-temperature deconf\/ined phase $\langle \Lx\rangle\neq 0$.
The ef\/fective action for the Polyakov loop dynamics is
\begin{gather}
e^{-\Seff[{\cal P}]}= \int {\cal D}U \delta\left({\cal
P}_{\mbx},\prod_{t=1}^{N_t} U_{t,\mbx;0}\right) \,e^{-S_{\rm
w}[U]} ,\qquad {\cal D}U=\prod_{\rm links}d\mu_{\rm
Haar}(U_{x;\mu}),\label{pleffaction}
\end{gather}
with gauge f\/ield action $S_{\rm w}$. In this formula the group
valued f\/ield ${\cal P}_\mbx$ is prescribed and the
del\-ta-distribution enforces the constraints \refs{constr}. In
the simulations we used the Wilson action for the gauge f\/ields.
Gauge invariance of the action $S_{\rm w}[U]$ and measure ${\cal
D}U$ implies $\Seff[{\cal P}]=\Seff[L]$. In addition there is the
global center symmetry, under which all ${\cal P}_\mbx$ are
multiplied by the same
\begin{wrapfigure}[16]{l}{7cm}
\centerline{\includegraphics{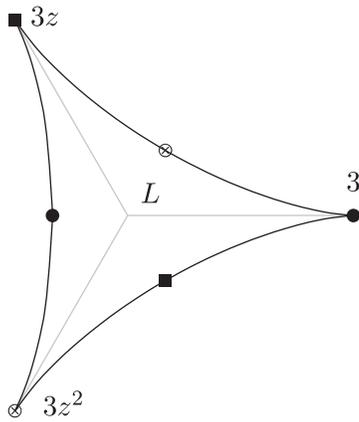}}
\caption{\label{fig:fudomain} Fundamental domain of $L$.}
\end{wrapfigure}
center element of the group. For $SU(N)$ the center consists of
the $N$ roots of unity, multiplied by the identity matrix. Hence
for $SU(N)$ theories we have
\begin{gather*}
\Seff[L]=\Seff[z\cdot L],\qquad z^N=1.
\end{gather*}
In Fig.~\ref{fig:fudomain} on the left we plotted the domain of
the traced Polyakov loop variable for $SU(3)$. The values of $L$
at the three center elements are $3$, $3z$ and $3z^ 2$ with
$z=e^{2\pi i/3}$. They form the edges of the triangle. What is
needed is a good ansatz for the ef\/fective action $\Seff$ in
\refs{pleffaction}. To this aim we calculated the leading terms in
the strong coupling expansion for $\Seff$ in gluodynamics
\cite{wozar,buss}. As expected one f\/inds a character expansion
with nearest neighbor interactions,
\begin{equation*}
 \Seff = \lambda_{10} S_{10} + \lambda_{20} S_{20} + \lambda_{11}
 S_{11} + \lambda_{21} S_{21} + \rho_1 V_1 + O(\beta^{3N_t}) ,
\end{equation*}
where $S_{pq}$ depends on the character $\chi_{pq}$ belonging to
the representation $(p,q)$ of $SU(3)$. Expressing the characters
as function of the fundamental characters $\chi_{10}=L$ and
$\chi_{01}=L^*$ the dif\/ferent center-symmetric contributions
have the form
\begin{gather*}
  S_{10} = \sum_{\langle \mbx \mby \rangle} \left( \Lx \Ly^* +
  \text{h.c.} \right) ,  \\
  S_{20} = \sum_{\langle \mbx \mby \rangle} \left( \Lx^2 \Ly^{*2} - \Lx^2
  \Ly - \Lx^* \Ly^{*2} + \Lx^* \Ly + \text{h.c.} \right) , 
  \\ S_{11} = \sum_{\langle \mbx \mby \rangle} \left( |\Lx|^2 |\Ly|^2 -
  |\Lx|^2 - |\Ly|^2 + 1 \right)  ,\\
  S_{21} = \sum_{\langle \mbx \mby \rangle} \left( \Lx^2 \Ly  +
  \Ly^2 \Lx - 2\Lx^* \Ly + \text{h.c.} \right)  ,  \qquad
  V_1 = \sum_{\mbx} \left( |\Lx|^2 - 1 \right).
\end{gather*}
The target space for $L$ is the fundamental domain inside the
triangle depicted above. The functional measure is not the product
of Lebesgue measures but the product of reduced Haar measures on
the lattice sites.

In the following we consider the two-coupling model
\begin{equation}
\Seff= (\lam_{10}-2\lam_{21})\sum\left(L_\mbx
L^*_\mby+\hbox{h.c.}\right)
+\lam_{21}\sum\left(L_\mbx^2L_\mby+L_\mby^2
L_\mbx+\hbox{h.c.}\right)\label{seff1}
\end{equation}
which contains the leading order contribution in the strong
coupling expansion. For vanishing~$\lam_{21}$ it reduces to the
Polonyi--Szlachanyi model \cite{polonyi}.

\section{Gluodynamics and Potts-model}
There is a direct relation between the ef\/fective Polyakov loop
dynamics and the $3$-state Potts model: If we freeze the Polyakov
loops to the center of the group,
\begin{equation*}
{\cal P}_\mbx\longrightarrow z_x\id\in
\hbox{center}\big(SU(3)\big)\Longleftrightarrow
 \theta_\mbx\in\left\{0,\frac{2\pi}{3},\frac{4\pi}{3}\right\}
\end{equation*}
then the ef\/fective action \refs{seff1} reduces to the
Potts-Hamiltonian \refs{ham} with $J=18(\lam_{10}+4\lam_{21})$.
The same reduction happens for all center-symmetric ef\/fective
actions with nearest neighbor interactions. Only the relation
between the couplings $\lam_{pq}$ and $J$ is modif\/ied.

For the gauge group $SU(2)$ the f\/inite temperature phase
transition is second order and the critical exponents agree with
those of the $2$-state Potts spin model which is just the
ubiquitous Ising model. The following numbers are due to Engels et
al.~\cite{engels}
\begin{center}
\begin{tabular}{|c|ccc|}\hline
&$\beta/\nu$& $\gamma/\nu$&$\nu$\\ \hline
$4d$ SU(2)&$0.525$& $1.944$&$0.630$\\
$3d$ Ising&$0.518$& $1.970$&$0.629$ \\ \hline
\end{tabular}
\end{center}
and support the celebrated Svetitsky--Yaf\/fe conjecture
\cite{Yaffe:1982qf}\footnote{For the relations between
$3$-dimensional gauge theories at the deconf\/ining point and
$2$-dimensional Potts-models, the so-called gauge-CFT
correspondence, we refer to the recent paper \cite{gliozzi}.}. For
the gauge group $SU(3)$ the f\/inite temperature phase transition
is f\/irst order and we cannot compare critical exponents. But the
ef\/fective theory \refs{seff1} shows a second order transition
from the symmetric to a antiferromagnetic phase and we can compare
critical exponents at this transition with those of the same
transition in the $3$-state Potts spin model.

\begin{figure}[t]
\centerline{\includegraphics[width=10cm]{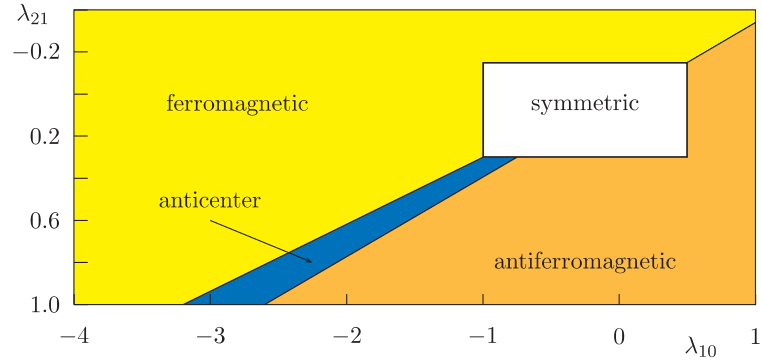}}
\caption{\label{fig:clPhaseDiagram} Classical phase diagram.}
\end{figure}

In a f\/irst step we study the `classical phases' of the Polyakov
loop model with action \refs{seff1}. The classical analysis, where
one minimizes the `classical action' $\Seff$, shows a
ferromagnetic, antiferromagnetic and anticenter phase. The
classical phase diagram is depicted in
Fig.~\ref{fig:clPhaseDiagram}. For small couplings the quantum
f\/luctuations will disorder the system and entropy will dominate
energy. Thus we expect a symmetric phase near the origin in the
$(\lam_{10},\lam_{21})$-plane and such a~phase was inserted by
hand in the diagram. There exists one unexpected phase which
cannot  exist for Potts spin models. It is an ordered phase for
which the order parameter is near to the points opposite to the
center elements, hence we call it anticenter phase. They are
marked in the fundamental domain in Fig.~\ref{fig:fudomain}. We
shall see that the classical analysis yields the qualitatively
correct phase diagram.

\subsection[Modified mean field approximation]{Modif\/ied mean f\/ield approximation}
In a second step we calculate the ef\/fective potential for the
Polyakov loop model with action~\refs{seff1} in the mean f\/ield
approximation. Here we are not concerned with the relevance of
this model for f\/inite temperature gluodynamics and just
consider~\refs{seff1} as the classical action of a f\/ield
theoretical extension of the Potts spin models. We use the
variational characterization~\refs{effaction} for the  ef\/fective
action where we must minimize with respect to probability measures
on  the space of f\/ield conf\/iguration
 $\{{\cal P}_\mbx\vert\mbx\in\Lambda\}$ with f\/ixed expectation
values $\langle \chi_{pq}\rangle$  of all characters
$\chi_{pq}({\cal P}_\mbx)$ showing up in the Polyakov loop action.
As outlined in Section~\ref{sectpotts}, in the mean f\/ield
approximation we assume the measures to have product form,
\begin{equation*}
{\cal D P}\longrightarrow \prod_\mbx d\mu_{\rm red}({\cal
P}_\mbx)\,
p_\mbx \left({\cal P}_\mbx\right).
\end{equation*}
For further details the reader is referred to our earlier
paper~\cite{wipfsu2}. Here $\mu_{\rm red}$ is the reduced Haar
measure of $SU(3)$. Since we expect an antiferromagnetic phase we
only assume translational invariance on the sublattices in the
decomposition~\refs{prodstate2}. This way one arrives at a
non-trivial variational problem on two sites.

We illustrate the procedure with the simple Polyakov loop model
studied in \cite{polonyi},
\begin{equation}
\Seff=\lam S_{10}= \lam\sum\left(L_\mbx
L^*_\mby+\hbox{h.c}\right).\label{minimal}
\end{equation}
To enforce the two constraints $\langle L_\mbx\rangle=L_i$ for
$\mbx\in\Lambda_i$ one introduces two Lagrangian multipliers. For
the minimal model \refs{minimal} one arrives at the following mean
f\/ield ef\/fective potential
\begin{gather*}
2u_{\rm mf}(L_1,L_1^*,L_2,L_2^*)= -d\lambda\vert L_1-L_2\vert^2
+\sum v_{\rm mf}(L_i,L_i^*)\\
\hbox{with}\qquad v_{\rm mf}(L,L^*)=d\lambda\vert
L\vert^2+\gamma_0(L,L^*).
\end{gather*}

\newpage

\begin{figure}[th]
\centerline{\includegraphics[width=10cm]{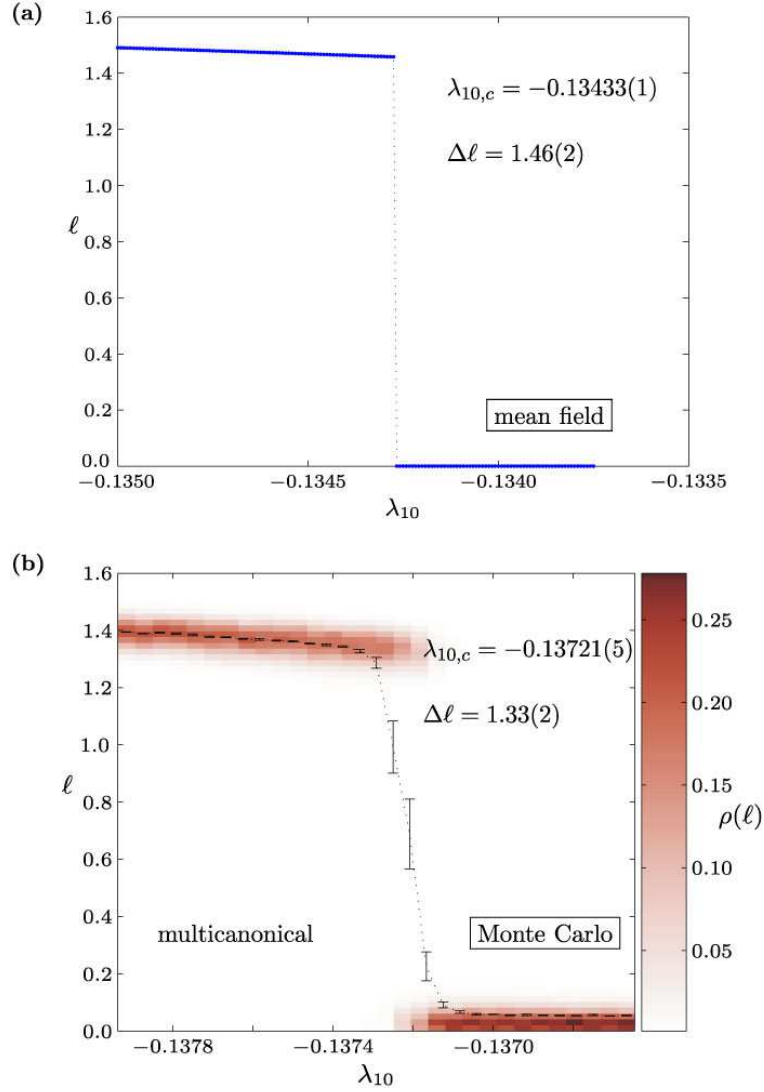}}
\caption{\label{fig:ferro}The order parameter $\ell$ for (a) mean
f\/ield approximation and (b) Monte Carlo simulation.}
\end{figure}

\noindent Here $\gam_0$ is the Legendre transform of
\begin{equation*}
w_0(j,j^*)=\log z_0(j,j^*),\qquad z_0(j,j^*) ={\int d\mu_{\rm
red}\exp\left(jL+j^ *L^*\right).}
\end{equation*}
The last integral has an expansion in terms of modif\/ied Bessel
functions \cite{bessel,uhlmann},
\begin{equation*}
z_0(j,j^*)= \sum_{n\in\,\Z}e^{inN\alpha}\, \det \begin{pmatrix}
I_n&I_{n+1}&I_{n+2}\cr I_{n-1}&I_n&I_{n+1}\cr I_{n-2}&I_{n-1}&I_n
\end{pmatrix}
\left(2\vert j\vert\right),\qquad j=\vert j\vert e^{i\alpha}.
\end{equation*}
As order parameters discriminating the symmetric, ferromagnetic
and antiferromagnetic phases we take \cite{wang}
\begin{equation}
L=\frac{1}{2}(L_1+L_2),\qquad M=\frac{1}{2}(L_1-L_2),\qquad
\ell=\vert L\vert,\qquad m=\vert M\vert.\label{orderp}
\end{equation}
Fig.~\ref{fig:ferro} shows the value of the order
parameter $\ell$ as function of the coup\-ling $\lam_{10}$ near
the phase transition point from the symmetric to the ferromagnetic
phase. The value of the critical coupling and the jump $\Delta
\ell$ of the order parameter in the mean f\/ield approximation and
simulations agree astonishingly well. In Fig.~\ref{fig:ferro}b the
order parameter $\ell$ for the ferromagnetic phase is plotted
against its probability distribution given by the shaded area. In
the vicinity of the f\/irst order transition we used
a~multicanonical algorithm on a $16^3$ lattice to calculate the
critical coupling to very high precision. For further details on
algorithmic aspects I refer to our recent paper \cite{wozar}.
Since the model has the same symmetries and dynamical degrees as
the 3-state Potts model in  3 dimensions we expect the model to
have a second order transition from the symmetric to the
antiferromagnetic phase. To study this transition we calculated
the order parameter $m$ in~\refs{orderp} in the modif\/ied mean
f\/ield approximation and with Monte Carlo simulations. The order
parameter $m$ is sensitive to the transition in question.

\begin{figure}[t]
\centerline{\includegraphics[width=10cm]{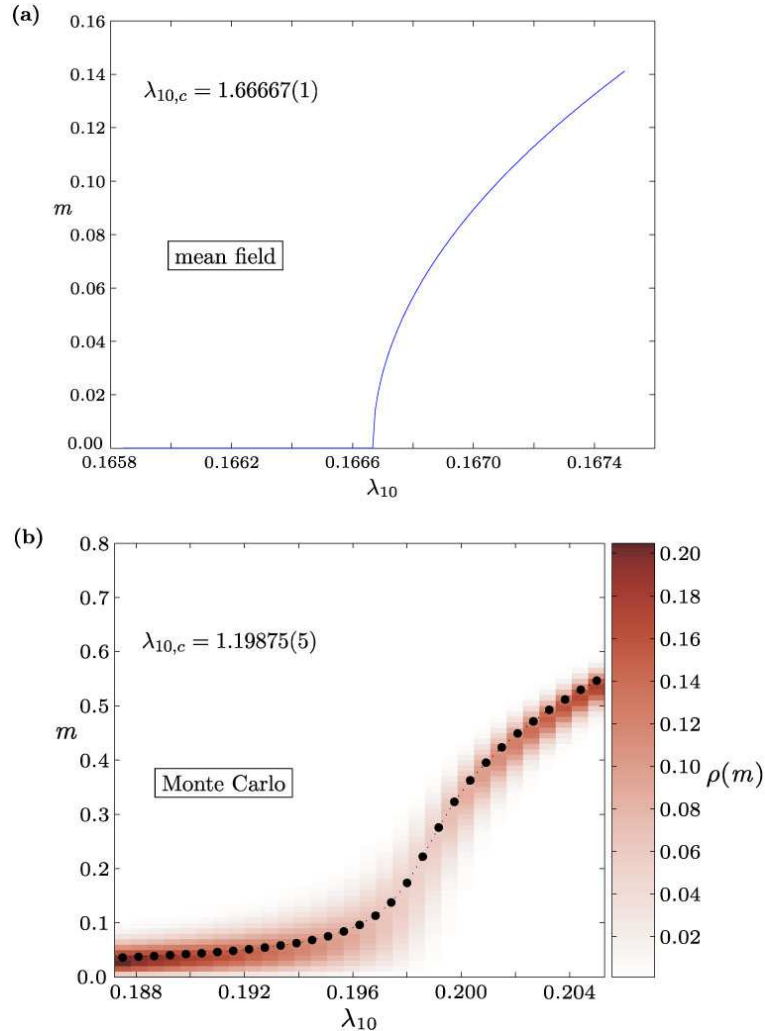}}

\vspace{-2mm} \caption{\label{fig:aferro}The order parameter $m$
for (a) mean f\/ield approximation and (b) Monte Carlo
simulation.}
\vspace{-2mm}

\end{figure}

Again the expectation value and probability distribution of $m$
near the transition is plotted in Fig.~\ref{fig:aferro}. To get a
clear signal we have chosen a large lattice with $28^ 3$ sites and
evaluated $5\times 10^5$ sweeps. The mean f\/ield approximation
and Monte Carlo simulations together with the calculation of
Binder cumulants in \cite{wozar} indicate that the transition to
the antiferromagnetic phase is second order. We plan to use a
f\/inite size scaling method to conf\/irm this result in our
upcoming work.

With the cumulant method we have calculated the critical exponents
$\gamma$ and $\nu$ and compared our results with the same
exponents for the $3$-state Potts model at the second order
transition from the symmetric to the antiferromagnetic phase
\cite{wang}. Within error bars the critical exponents are the
same.
 \begin{center}
\begin{tabular}{|c|cc|}\hline
exponent&3-state Potts& minimal $S_{\rm eff}$\\ \hline
$\nu$ & $0.664(4)$ & $0.68(2)$\\
$\gamma/\nu$ & $1.973(9)$ & $1.96(2)$\\ \hline
\end{tabular}
\end{center}
This is how the conjecture relating f\/inite temperature
gluodynamics with spin models is at work: The Polyakov loop
dynamics is ef\/fectively described by generalized Potts models
with the fundamental domain as target space, for $SU(3)$ it is the
triangularly shaped region in Fig.~\ref{fig:fudomain}. The f\/irst
order transition of gluodynamics is modelled by the transition
from the symmetric to the ferromagnetic phase in the generalized
Potts models. These generalized Potts models are in the same
universality class as the ordinary Potts spin models; they have
the same critical exponents at the `unphysical' second order
transition from the symmetric to the antiferromagnetic phase.

It is astonishing how good the mean f\/ield approximation is. The
reason is probably, that the upper critical dimension of the
(generalized) $3$-state Potts model is $3$ and not $4$ as one
might expect. This is explained by the fact, that the models are
embedded in systems with tricritical points, see below, and for
such systems the upper critical dimension is
reduced~\cite{lawrie,lawrie1}.

\section[Simulating the effective theories]{Simulating the ef\/fective theories}
We have undertaken an extensive and expensive scan to calculate
histograms in the coupling constant plane
$(\lambda_{10},\lambda_{21})$ of the model~\refs{seff1}. Away from
the transition lines we used a standard Metropolis algorithm
giving results within 5 percent accuracy. Near f\/irst order
transitions we simulated with a multicanonical algorithm on
lattices with up to $20^3$ lattice sites. Most demanding have been
the simulations near second order transitions. For that we
developed a~new \emph{cluster algorithm} \cite{wozar} which
improved the auto-correlation times by two orders of magnitude on
larger lattices. We found a rich phase structure with $4$
dif\/ferent phases with second and f\/irst order transitions and
tricritical points. As for the minimal model the Monte Carlo
simulations are in good and sometimes very good agreement with the
mean f\/ield analysis.

\begin{figure}[h]
\centerline{\includegraphics[width=15cm]{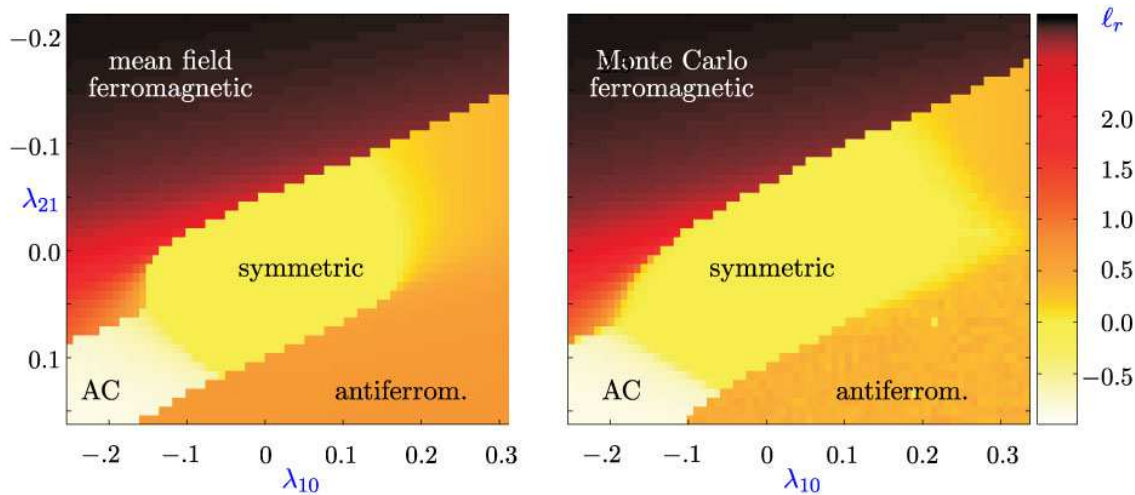}}
\caption{\label{fig:phaseDiagrams} Phase structure for MF
approximation and MC simulation.}
\end{figure}

\begin{figure}[H]
\centerline{\includegraphics[width=9.5cm]{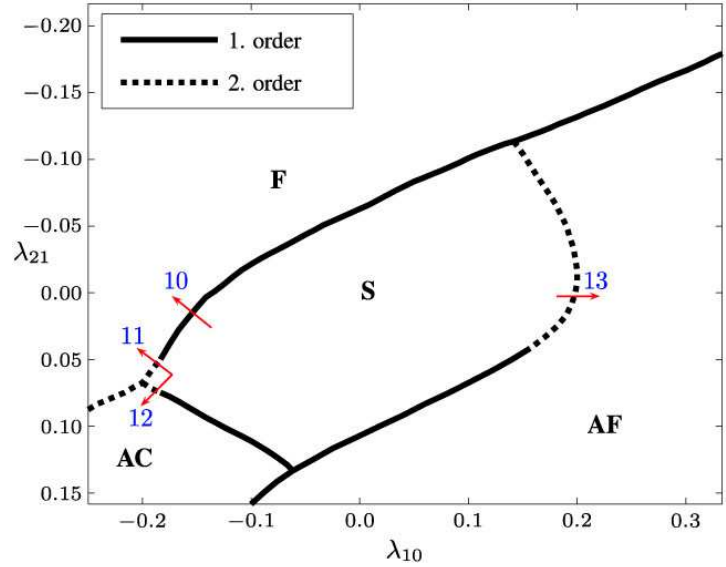}}

\vspace{-2mm}
\caption{\label{fig:phasePortrait} Phase portrait of the model
\refs{seff1}. Histograms for the transitions are marked with
arrows and f\/igure numbers.}
\end{figure}

Fig.~\ref{fig:phaseDiagrams} shows the phase structure in the
generalized MF approximations and the corresponding results of our
extended MC simulations. The results of the simulations are
summarized in the following phase portrait (Fig.
\ref{fig:phasePortrait}), in which we have indicated the order of
the various transitions. The calculations were done on our Linux
cluster with the powerful jenLaTT package. In $3000$ CPU hours we
calculated $8000$ histograms in coupling constant space.

\begin{figure}[th]
\centerline{\includegraphics{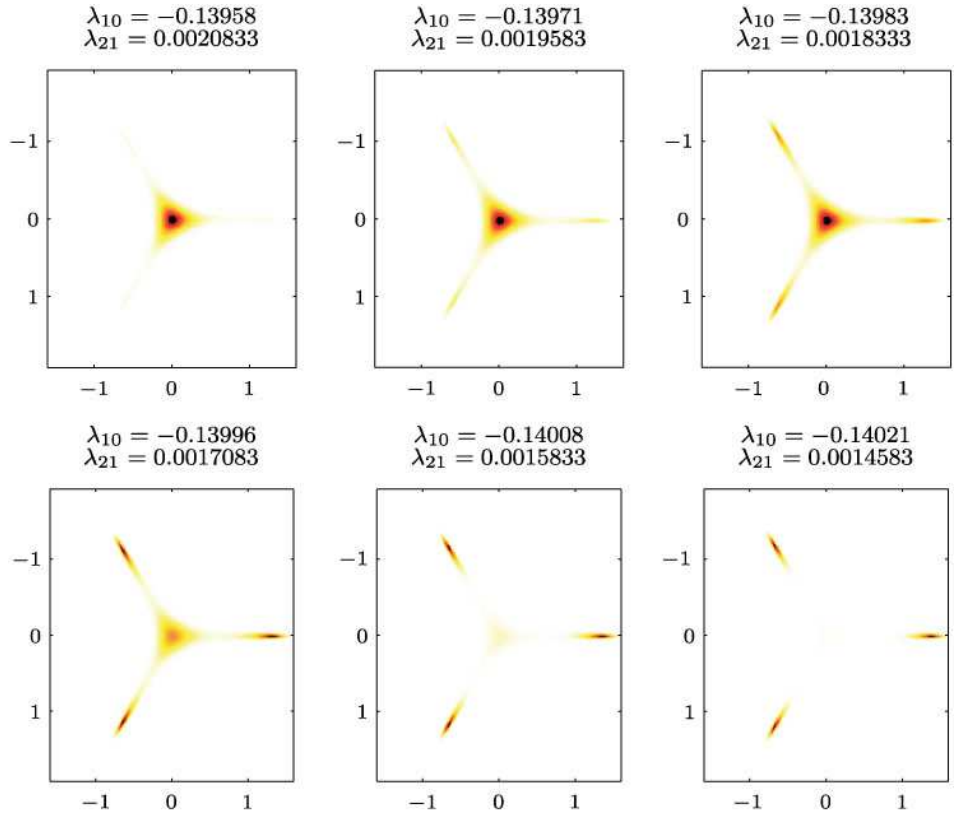}}

\vspace{-2mm}
\caption{\label{fig:histos1} Histograms of $L$ which show a clear
signal for the f\/irst order transition from the symmetric to the
ferromagnetic phase.}\vspace{-2mm}
\end{figure}

\begin{figure}[t]
\centerline{\includegraphics[width=13.0cm]{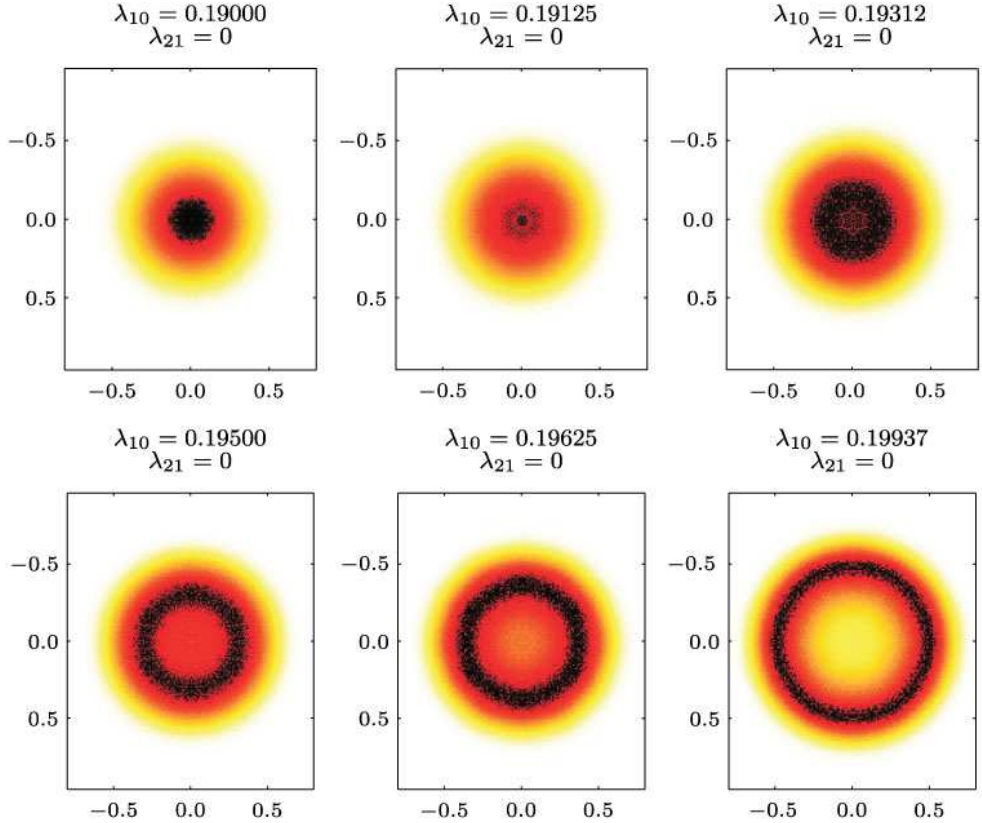}}
\vspace{-1mm}
\caption{\label{fig:histos4} Histograms of $M$ for the second
order transition from the symmetric to the antiferromagnetic
phase.}
\end{figure}

For each marked transition in Fig.~\ref{fig:phasePortrait} a set
of $6$ histograms is displayed. These histograms (and many others,
see \cite{wozar}) have been used to localize the phase transition
lines and to investigate the nature of the transitions. The
results are summarized in the phase portrait on page
\pageref{fig:phasePortrait}. The critical exponents $\nu$ and
$\gamma$ given above have been determined at the second order
transition indicated with arrow \ref{fig:histos4} in the portrait.

\begin{figure}[tp]
\centerline{\includegraphics[width=13cm]{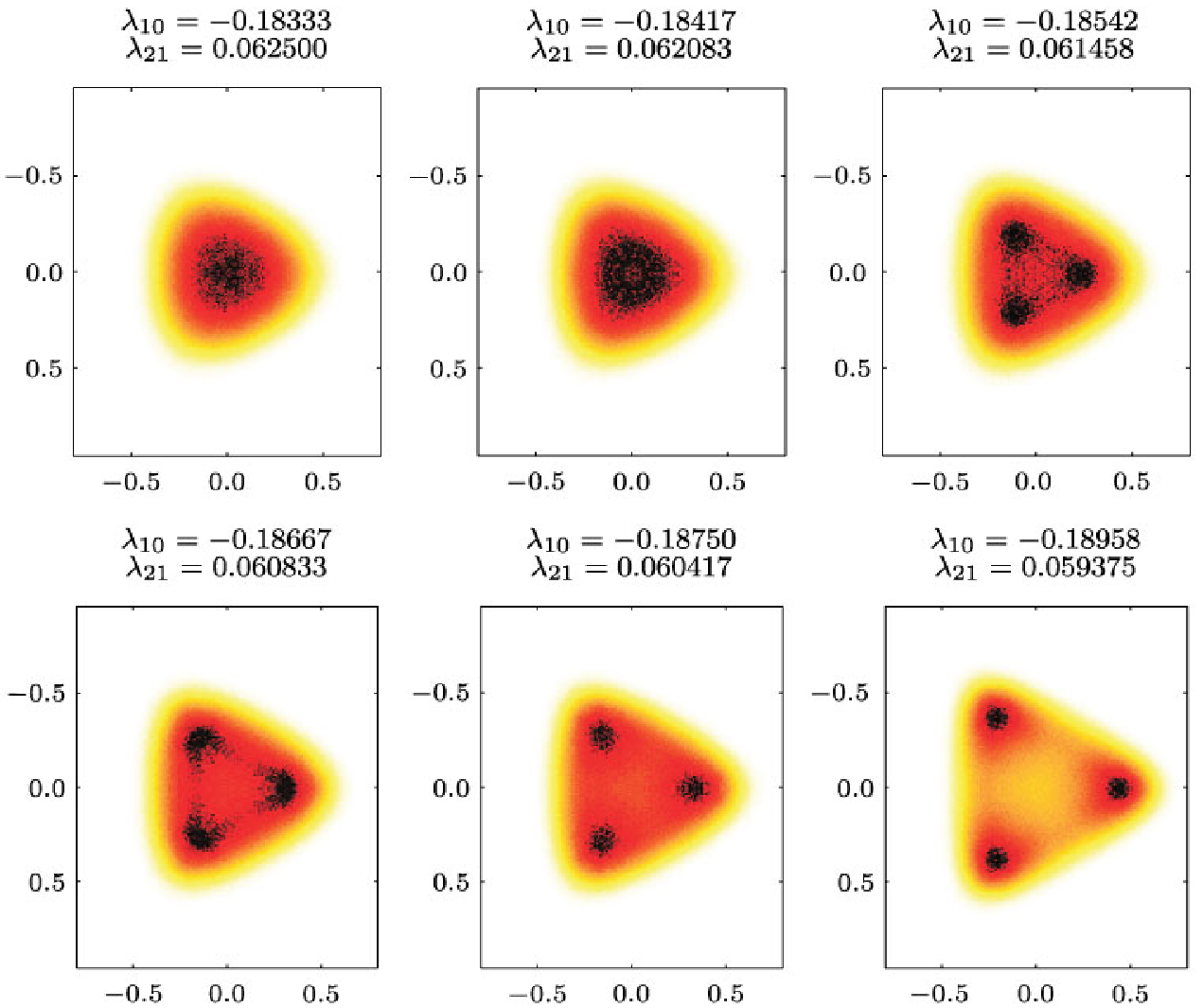}} 
\vspace{-2mm}
\caption{Histograms of
$L$ for the continuous transition from the symmetric to the
ferromagnetic phase.\label{fig:histos2}}\vspace{2mm}

\centerline{\includegraphics[width=13cm]{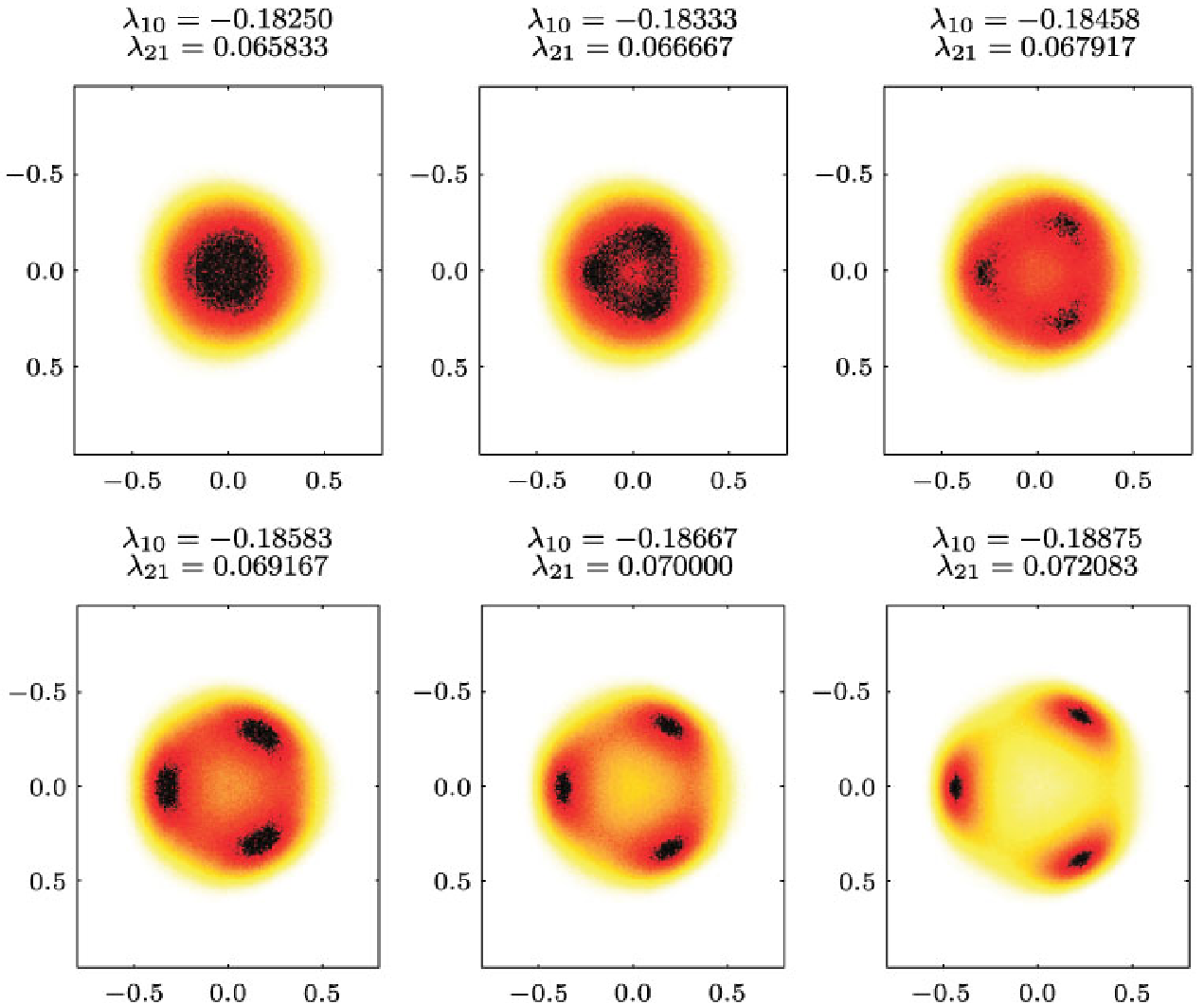}}
\vspace{-2mm}
\caption{\label{fig:histos3} Histograms of $L$ for the continuous
transition from the symmetric to the anticenter
phase.}\vspace{-2mm}
\end{figure}

\section{Conclusion}
The strong coupling expansion results in a character expansion for
the Polyakov-loop dynamics. The leading terms are center symmetric
nearest neighbor interactions containing  the characters of the
smallest representations of the gauge group. We have performed an
extensive modif\/ied mean f\/ield analysis which includes
antiferromagnetic states without translational invariance on the
whole lattice. A new and ef\/f\/icient cluster algorithm has been
developed and applied to study the second order transitions from
the symmetric to the antiferromagnetic phase. The autocorrelation
times were improved by $2$ orders of magnitude. We discovered an
unexpectedly rich phase structure of the simple $2$-coupling
Polyakov loop model \refs{seff1}. This model is in the same
universality class as the $3$-state Potts spin model. The mean
f\/ield results are surprisingly accurate that seems to indicate
that the upper critical dimension of the generalized Potts models
is $3$. This is attributed to the existence of tricritical points
\cite{lawrie,lawrie1}.

To relate our results to gluodynamics we have to calculate the
ef\/fective couplings $\lam_{pq}$ gover\-ning Polyakov-loop dynamics
as functions of the Wilson coupling $\beta$ in gluodynamics. We
have done this successfully for $SU(2)$ gauge theory with inverse
Monte Carlo techniques \cite{wipfsu2,imcsu2} and plan to publish
our results for $SU(3)$ very soon \cite{imcsu3}. For the inverse
Monte Carlo simulations to work one needs simple geometric
Schwinger Dyson equations for the Polyakov loop dynamics. Such
equations have been derived very recently in \cite{uhlmann}. It
would be interesting to see whether the antiferromagnetic phase of
the Polyakov loop models plays any role at all for gluodynamics.
In the present paper it was needed to show that certain critical
exponents of the Polyakov loop models are the same as of the $q=3$
Potts spin model. Finally one would like to include heavy fermions
in the ef\/fective Polyakov-loop dynamics \cite{Kim:2005ck}. To
that end one needs to add center symmetry breaking terms to the
ef\/fective actions studied in the present paper. This will lead
to a proliferation of additional terms in the ef\/fective action
which renders a systematic study more dif\/f\/icult as compared to
pure gluodynamics.

\subsection*{Acknowledgements} Andreas Wipf would like to thank
the local organizing committee of the  \emph{O'Raifeartaigh
Symposium on Non-Perturbative and Symmetry Methods in Field
Theory} for organizing such a stimulating and pleasant meeting on
the hills of Budapest to commemorate Lochlainn O'Raifeartaigh and
his important contributions to physics. This contribution is very
much in the tradition of his research on the role of symmetries
and ef\/fective potentials in f\/ield theory.

\pdfbookmark[1]{References}{ref}
\LastPageEnding

\end{document}